\documentclass[fleqn,usenatbib]{mnras}

\usepackage{newtxtext,newtxmath}
\usepackage{color}

\usepackage[T1]{fontenc}
\usepackage{siunitx}

\usepackage{newtxtext,newtxmath}
\usepackage{booktabs}
\usepackage{lscape}
\usepackage{graphicx}	% Including figure files
\usepackage{amsmath}	% Advanced maths commands
\usepackage{multirow} 
\title{Observing white dwarf tidal stripping with TianQin gravitational wave observatory}

\author[C. Q. Chang et al.]{Chang-Qing Ye$^{1, 2}$,
 Jin-Hong Chen$^{1, 3}$,
Jian-dong Zhang$^{1, 2}$,
Hui-Min Fan$^{1, 2}$\thanks{Corresponding author: fanhm3@mail.sysu.edu.cn},
Yi-Ming Hu$^{1, 2}$
\\
$^1$ School of Physics and Astronomy, Sun Yat-sen University (Zhuhai Campus), Zhuhai 519082, China. \\
$^2$ MOE Key Laboratory of TianQin Mission, TianQin Research Center for
Gravitational Physics, Frontiers Science Center for TianQin, \\
 Gravitational Wave Research Center of CNSA, Sun Yat-sen University (Zhuhai Campus), Zhuhai 519082, China. \\
$^3$ Department of Physics, University of Hong Kong, Pokfulam Road, Hong Kong, China.
}

\begin{document}
\label{firstpage}
\pagerange{\pageref{firstpage}--\pageref{lastpage}}
\maketitle

\begin{abstract}
Recently discovered regular X-ray bursts known as quasi-periodic eruptions have a proposed model that suggests a tidal stripping white dwarf  inspiralling into the galaxy's central black hole on an eccentric orbit. According to this model, the interaction of the stripping white dwarf with the central black hole would also emit gravitational wave signals, their detection can help explore the formation mechanism of quasi-periodic eruptions and facilitate multi-messenger observations. In this paper, we investigated the horizon distance of TianQin on this type of gravitation wave signal and find it can be set to 200Mpc. We also find that those stripping white dwarf model sources with central black hole mass within $10^4\sim10^{5.5}M_\odot$ are more likely to be detected by TianQin. We assessed the parameter estimation precision of TianQin on those stripping white dwarf model sources. Our result shows that, even in the worst case, TianQin can determine the central black hole mass, the white dwarf mass, the central black hole spin, and the orbital initial eccentricity with a precision of $10^{-2}$. In the optimistic case, TianQin can determine the central black hole mass and the white dwarf mass with a precision of $10^{-7}$, determine the central black hole spin with a precision of $10^{-5}$, and determine the orbital initial eccentricity with a precision of $10^{-8}$. Moreover, TianQin can determine the luminosity distance with a precision of  $10^{-1}$ and determine the sky localization with a precision of $10^{-2}\sim10$ $\rm deg^2$.
\end{abstract}

% (QPEs) are periodic X-ray burst events, with one theoretical model proposed as t
% events proposed as one theoretical model for quasi-periodic eruptions (QPEs),

%%%%%%%%%%%%%%%%%%%%%%%%%%%%%%%%%%%%%%%%%%%%%%%%%%

%%%%%%%%%%%%%%%%% BODY OF PAPER %%%%%%%%%%%%%%%%%%

\section{Introduction}

Currently, multiple new type X-ray burst  events with a recurrence time of  several hours are detected in the galaxy nucleus (\cite{Miniutti:2019fqr, Giustini:2020gex, Arcodia:2021tck, Chakraborty:2021lfh}), which are referred as quasi-periodic eruptions (QPEs).  To explain the characteristics of the QPEs, several theoretical models are proposed. One of them is the stripping white dwarf model showing that a low-mass white dwarf (WD) extreme-mass-ratio inspiral (EMRI) on a highly eccentric orbit (\cite{Zalamea:2010mv,King:2020jtd,Zhao:2021swe,Wang:2022irk}). In this model, the WD overflows its Roche lobe after approaching the tidal radius, and its exterior layer is tidally stripped during the pericentric encounters. The stripped mass accreted by the central BH produces electromagnetic radiation, accounting for the periodic X-ray burst signals. 

Apart from this model, there are also other predicted models including the radiation pressure accretion disk instability (\cite{Janiuk:2002nr,Merloni:2006pb,Janiuk:2011xg}), the gravitational self-lensing binary supermassive black hole (SMBH) model (\cite{Ingram:2021gar}), the star-disk collision model (\cite{Dai2010:qpe,Xian:2021qpe,Franchini:2023qpe}) and periodic close interactions between two  coplanar stellar EMRIs (\cite{Metzger:2021zia}).
By now, the QPE formation mechanism remains uncertain and is still a mystery. In the stripping white dwarf model, the interaction of the WD with the central BH  also generates GW signals. Detecting these GW signals can provide a new method for investigating astronomical information and is worth further exploration. The GW signals under the stripping white dwarf model (here, we simplify them as SWDs ) are expected to have mHz frequencies, which can be probed by TianQin and LISA.
%All of these models can not fully explain the observational characteristics of QPEs, which make the QPE formation mechanism still remain a mystery

%GW detection attract much interest since the first detection of GW150914 by LIGO and Virgo (\cite{LIGOScientific:2016aoc}). To date, LIGO and Virgo have detected tens of GW events (\cite{LIGOScientific:2021djp}). Due to limited seismic noise and short detector arm length, LIGO and Virgo are mainly focused on low-mass sources. To overcome these limitations and achieve the detection of  more massive sources, GW detectors put in space are proposed. 
TianQin and LISA are  proposed space-based GW observatories. According to the project design, LISA will have an armlength of $2.5\times10^6$km, with a mission lifetime of 4 years and aiming to detect GW sources in the frequency band of $10^{-5}-10^{-1}$Hz (\cite{Amaro-Seoane:2022rxf,Baker:2019nia,Karnesis:2022vdp}). Different from LISA, which put in the orbit lag behind the Earth about $20$ degrees, TianQin is a geocentric space-based GW observatory with armlength of $1.7\times10^5$km, aiming to detect GW sources in the frequency band of $10^{-4}-1$Hz (\cite{TianQin:2015yph,TianQin:2020hid}).   

So far, TianQin and LISA have published a series of works to study the science prospects of various sources, including Galactic ultra-compact binaries (\cite{Hu:2018yqb,Huang:2020rjf,Brown:2020uvh,Kremer:2017xrg,Korol:2017qcx}), coalescing massive black holes (\cite{Wang:2019ryf,Feng:2019wgq,Ruan:2021fxq,Shuman:2021ruh,Katz:2019qlu}), the mergers of intermediate-mass black holes (\cite{Fragione:2022ams, Torres-Orjuela:2023hfd}), the low-frequency inspirals of stellar-mass black holes (\cite{Liu:2020eko,Klein:2022rbf,Buscicchio:2021dph,Ewing:2020brd,Toubiana:2020cqv}), the EMRIs (\cite{Fan:2020zhy,Zhang:2022xuq,Wardell:2021fyy,Lynch:2021ogr,Isoyama:2021jjd,Vazquez-Aceves:2022dgi}) and the stochastic GW backgrounds (\cite{Liang:2021bde,Renzini:2022alw,LISACosmologyWorkingGroup:2022kbp,Boileau:2020rpg}). The detection of these sources has a lot of significance, such as testing general relativity (\cite{Zi:2021pdp,Shi:2019hqa}), gaining a glimpse into the source formation and evolution history (\cite{Fan:2022wio,Amaro-Seoane:2007osp}). 

Here, as mentioned earlier, the detection of  SWDs can provide an opportunity to explore the QPE formation mechanism. Besides,  SWDs are nice sources for the precision measurement of the Hubble constant. This is because the SWDs have corresponding electromagnetic signals. Thus, one can obtain the luminosity distance information from the GW observation and extract the redshift information from the electromagnetic counterpart, then perform the cosmological inference (\cite{Ye:2023fpb, Zhu:2021aat,Wang:2020dkc,Kyutoku:2016zxn,MacLeod:2007jd}).

%in either case each QPE is too weak to be resolvable by LISA
 
Previous works (\cite{Chen:2021ool,  Chen:2022oin}) indicate that the SWD signals from the five observed QPE events are too faint to be resolved by LISA and TianQin. In light of this, how far SWDs can be detected, and what parameters of those SWD sources raise a question for us. Furthermore, the WD will experience mass loss and tidal deformability during the evolution, 
the influence of these two effects on the SWD waveforms has not been explored.

In this paper, we aim to give a preliminary assessment of the detectability of TianQin on SWD signals. We explore the stripping white dwarf model source distribution and utilize an analytical kludge method with higher-order modes to obtain the waveform. We add mass loss and tidal deformability to the WD evolution.
 Based on the performance of the TianQin detector, we obtain the horizon distance to quantify the farthest distance at which SWDs can be detected. Additionally, we estimate the maximum detection distance of the corresponding electromagnetic signals by the Einstein Probe. We expect the combined analysis of these two types of signals would provide more insights into astronomical exploration. Finally, We assess TianQin's capability to estimate the SWD  source parameters, whose accurate estimation will provide useful information to the study of the relativistic universe.

%t is evident that the farther of the galaxies from which QPEs are detectable, the more possible of the QPE GW events are expected to be detected in the future. 

This paper is organized as follows: In Sec.\ref{eventDistri}, we describe the stripping white dwarf model source distribution. In Sec.\ref{wave}, we illustrate the waveform calculation method and the two effects that would influence SWD waveforms. In Sec.\ref{detector}, we describe the TianQin gravitational wave detector and its response to SWD signals. In Sec.\ref{result}, we present our methods and results. Finally, in Sec.\ref{conclusions}, we provide our conclusions and discussion.

\section{Stripping White Dwarf Model Source Distribution}\label{eventDistri}

In the vicinity of  the central BH of the galaxy, the stars are constantly subject to gravitational perturbations by other stars, which might cause a close encounter between a WD and the central BH (\cite{Amaro-Seoane:2012lgq}). The less bound WD with eccentricity $e \simeq 1$ could suffer  a single-passage (\cite{Rees1988TidalDO}).  On the contrary, the tightly bound WD with $e \lesssim 1$ could undergo a multiple-passage and form a very eccentric  EMRI event (\cite{MacLeod:2014mha, MacLeod:2015bpa}). In the latter case, the central BH produces a tidal field on the WD, deforming its shape and multipolar structure (\cite{Sesana:2008zc}). When the WD over-fill its Roche lobe radius, the tidal stripping begins.  Simulation shows that the WD loss mass only near the pericenter of the orbit, even with a very small orbital eccentricity (\cite{Zalamea:2010mv}). This type of event is considered as  one theoretical model of QPEs, which could also form through binary splitting (\cite{Hills1988HypervelocityAT,Wang:2022irk}).

%Thus the periodic inspiral of  the WD lead to a periodic electromagnetic signal and constitute the formation mechanism of QPEs.

%WD get tidal deformed by the central BH and begin loss mass before its orbit become unstable. 
%The QPE events usually have highly eccentric orbit and typically happened in the dwarf galaxies with intermediate central massive blacks holes(IMBHs). 

Considering a central BH of mass $M$ and an inspiralling WD of mass $\mu$, the tidal stripping (\cite{Zalamea:2010mv}) happened when the orbital pericenter radius $R_{\rm p}$ at a distance away from the central BH with the value of
\begin{equation}\label{r_strip}
R_{\rm p}=R_{\rm strip}=2\times{R_{\rm T} }=2\times{R}_* (M/\mu)^{1/3} ,
\end{equation}
where $R_{\rm strip}$ is the tidal stripping radius, $R_{\rm T}$ is the tidal disruption radius from where the WD gets fully disrupted, $R_*$ is the WD's radius and has the equation as (\cite{Chen:2022oin, Paczyski1983ModelsOX})
\begin{equation}
R_*=9\times10^8\Bigg[1-\Big(\frac{\mu}{M_{\rm ch}}\Big)^{4/3}\Bigg]^{1/2}\Big(\frac{\mu}{M_\odot}\Big)^{-1/3}\rm cm ,
\end{equation}
where $M_{\rm ch}\simeq1.44M_\odot$ is the Chandrasekhar mass. 

Besides undergoing tidal stripping, the WD inspiraling towards the central BH in a bound orbit will  become unstable after reaching the last stable orbit (LSO) and be accreted by the central BH directly. The location of the LSO has been explored in many works (\cite{Cutler1994GravitationalRR,Stein:2019buj,Glampedakis:2002ya}), where for Schwarzschild spacetime, the  LSO has the value of $R_{\rm LSO}=(6+2e)r_g/(1+e)$, which correspond to a minimum value of $4r_g$, for Kerr spacetime, the calculation for LSO is more complicated, corresponding to a minimum value of $R_{\rm LSO}=r_g$. 
To be more conservative, we choose the value of $4r_g$ as the LSO of the WD, as most of the WDs can't reach the minimum orbital value of $r_g$.  

Then, we distinguish three regions: 

%the central BH is  Schwarzschild, its minimum last stable orbit $R_{\rm LSO}$ has the value of $4r_g$, which can be derived from the coalescence radius $r_{\rm coal}=(6+2e)r_g/(1+e)$\cite{cutler1994gravitational}, where $e$ is the orbital eccentricity at the plunge. %We consider the central BHs have spins and choose a relatively conservative LSO value of $4r_{ g}$ as the LSO of QPEs, as most of the QPE events can't reach the minimum value of $r_g$.  

(i) $R_{\rm LSO}\geq{R}_{\rm strip}$, the WD reaches the unstable orbit first and plunges into the central BH directly before tidal stripping begins.

(ii) $R_{\rm T}<R_{\rm LSO}<R_{\rm strip}$, the WD begins tidal stripped and plunges into the central BH finally before gets fully disrupted.

(iii) $R_{\rm LSO}\leq{R}_{\rm T}$, the WD begin tidal stripping and get fully disrupted.

\begin{figure}
\centering
\includegraphics[width=\linewidth]{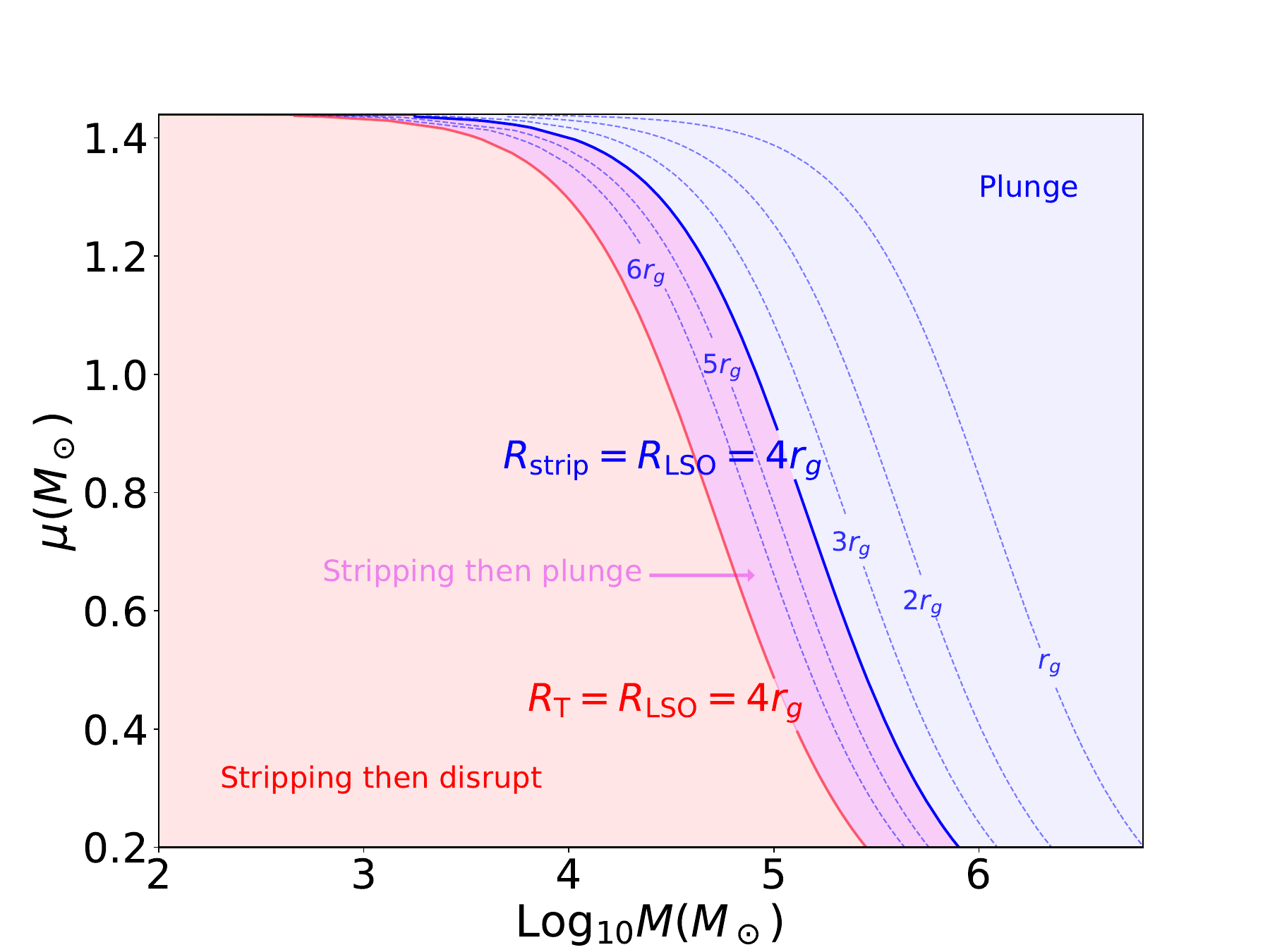}
\caption{The salmon and violet shadow region are where sources composed of the central BHs and the WDs can possibly form SWDs. The blue solid line is the distribution boundary corresponding to $R_{\rm strip}=4r_g$. Above the blue solid line (the light blue shadow region), the WDs will plunge into the central BHs directly. The red solid line is the distribution boundary corresponding to $R_{\rm T}=4r_g$. Below the red line (the salmon region), the WD will begin tidal stripping first and then get fully disrupted. Above the red line (the violet region), the WD will begin tidal stripping and then plunge into the central BH.  Distribution boundaries with $R_{\rm LSO}=6r_g, 5r_g, 3r_g, 2r_g, r_g$ are also plotted in blue dashed lines.}
\label{fig:eventdistribut}
\end{figure}

\noindent The tidal stripping events refer to the sources that satisfy conditions (ii) and (iii), as sources with condition (i) can only be observed through the GW signals and are not expected to have electromagnetic signals.

By now, the discovered WDs are known to have the lowest mass of $\simeq0.2M_\odot$ (\cite{Istrate:2014hea}). Here, we choose this value as the least massive WD mass. The upper WD mass limit is the Chandrasekhar mass, as the WD can no longer be supported by electron degeneracy pressure after its mass reaches this value. Hence, we have the WD mass range between  $0.2M_\odot-M_{\rm ch}$. We do not concern those SWD sources with central BH mass less than $10^2M_\odot$, as we do not expect they will exist in the galaxy center. Then, based on the equation (\ref{r_strip}) and the $R_{\rm LSO}$, the tidal stripping events satisfying conditions (ii) and (iii) are distributed as Fig.\ref{fig:eventdistribut}

In this figure, the SWD sources are distributed in the salmon and violet shadow region corresponding to $R_{\rm strip}=4r_g$, and the blue solid line marks the boundary upon which the WDs will plunge into the central BHs directly. The red solid line is the distribution boundary of $R_{\rm T}=4r_g$. Below the red line (the salmon region), the WD will begin tidal stripping first and then get fully disrupted. Above the red line (the violet region), the WD will begin tidal stripping and then plunge into the central BH. This figure also includes SWD distribution boundaries corresponding to $R_{\rm LSO}=6r_g, 5r_g, 3r_g, 2r_g, r_g$ with blue dashed lines.  From this figure, we can find that the SWDs will typically happen in dwarf galaxies with intermediate massive black holes (IMBHs).  Thus, surveys of these signals will also provide an opportunity to study the central BH with lower mass functions.

%To find the QPE parameter space,   we choose the WD mass ranged between $0.1M_\odot-M_{\rm ch}$, where $0.1M_\odot$ is the least massive WD mass, $M_{\rm ch}$ is the upper WD mass limit.

%for massive black hole(MBH), the tidal stripping radius is smaller than the last stable orbit, thus the WD will directly plunge into the MBH. The WD usually begin tidal stripped if its pericenter radius $R_{\rm p} \lesssim 2R_{\rm T}$, and get fully disrupted when $ R_{\rm p} \lesssim R_{\rm T}$,  where $R_{\rm T}$ is the tidal radius and is given by

%where $M$, $\mu$ and $R_*$ are the BH mass, WD's mass and radius, respectively. $R_*$ has the formula as

% a critical value $\beta_{\rm d} \simeq 1$ , 
%In fact, it is noteworthy to concern the relation between $R_{\rm p}$ and $R_{\rm T}$.
Defining a convenient impact factor as  $\beta \equiv R_{\rm T}/R_{\rm p}$, the critical value of $\beta_{\rm d}$ which separates the region where the WD get fully disrupted depends on the specific structure of the disrupted star. 
\cite{Ryu:2020huz} and \cite{Law-Smith:2020zkq} explored the tidal disruptions of the main-sequence stars and found $\beta_{\rm d} \simeq 0.9$ -- $2$. \cite{Guillochon2012HYDRODYNAMICALST}
found that $\beta_{\rm d} \simeq 0.9$ and $1.8$ for the polytropic stars with $\gamma = 5/3$ and $4/3$, respectively. For the least massive WD, its degenerate electrons are non-relativistic and the density profile is approximate to the $\gamma = 5/3$ polytropic star. On the contrary, for the most massive WD, its degenerate electrons are highly relativistic and the density profile is approximate to the $\gamma = 4/3$ polytropic star. However, for most of the WDs with a mass within that range, a single polytrope cannot depict the density profile 
(\cite{chandrasekhar1943dynamical}). 
Here, we assume $\beta_{\rm d} = 1$ that is between $0.9$ and $1.8$ for $\gamma = 5/3$ and $4/3$ polytropic stars, respectively. 
 In fact, we end the SWDs  at $\beta_{\rm d} = 0.7$ for some reasons, the explanation can be found in below Sec.\ref{MassLoss}.

%To our knowledge, the exact value of $\beta_{\rm d}$ has not been quantitatively explored for the WD, which merits a future investigation.   
% ($M_* \sim 1.4\ M_{\odot}$), ($M_* \sim 0.1\ M_{\odot}$)

 We assume the SWD sources have an initial eccentricity of $e_0= 0.9$ as the observed QPE events usually have a highly eccentric orbit. Additionally, We consider the central BHs in these systems to be Kerr BHs, which have spin.

\section{Waveform}\label{wave}

The waveform calculation for SWD has a similar method as EMRI. For EMRI, there are analytic kludge model (AK) (\cite{Barack:2003fp}), numerical kludge model (NK) (\cite{Babak:2006uv}), augmented analytic kludge model (AAK) (\cite{Chua:2017ujo}) and a recently developed method called FastEMRIwaveforms (FEW) (\cite{Katz:2021yft}) to obtain their waveform.  
The comparison of these theories shows that all of them get the main characteristic of the EMRI waveform, while AK has an advantage in terms of computing time, NK is physically more self-consistent, AAK and FEW have both advantages in computing time and waveform accuracy. Due to limitations from the algorithms, more accurate waveforms like AAK and FEW can not produce physical waveforms for certain values of the mass ratio and the semimajor axis. In this work, we adopt the simplified and computationally inexpensive method AK to obtain the SWD waveform.

% Therefore, we adopt the AK waveform throughout the calculation.

The AK waveform (\cite{Barack:2003fp}) is calculated using the quadrupole formula, with post-Newtonian equations including radiation reaction, pericenter precession, and Lense-Thirring precession to describe the WD orbital evolution. The quadrupole approximation waveform has the equation as
\begin{equation}
h_{ij}=(2/D_L)(P_{ik}P_{jl}-\frac{1}{2}P_{ij}P_{kl})\ddot{I}^{kl},
\end{equation}
where $P_{ij}\equiv\eta_{ij}-\hat{n}_i\hat{n}_j$ is the projection operator, $\hat{n}$ is the traveling direction $(\theta_S, \phi_S)$ of the SWD waveform, $D_L$ is the source luminosity distance, $I^{ij}$ is the inertia tensor of the WD.

The WD orbit is planar, $I^{ij}$ can be expressed as a sum of harmonics of the orbital frequency $\nu$: $I^{ij}=\sum_kI^{ij}_k$, where $I^{ij}_k$ is a function related with the first kind of Bessel functions $J_n$, $\Phi$, $e$, $\nu$,  $M$ and $\mu$, detailed expression can be found in \cite{Peters1963GravitationalRF} and \cite{Barack:2003fp}. Among these parameters, $\Phi$ is the mean anomaly of the WD's orbit, $e$ is the orbital eccentricity, $M$ is the central BH mass, and $\mu$ is the WD mass. As SWD has a quite eccentric orbit, more energy will be distributed into the higher-order mode. Here, we apply $k=100$ to obtain the SWD waveform. 
Knowing the pericenter precession angle $\gamma$, one can re-express the quantity $I^{ij}$ in the $\hat{L}$-based coordinate,
\begin{equation}
(\hat{x},\hat{y},\hat{z})_{ \hat{L}}:=\Bigg(\frac{(\hat{n}\cdot\hat{L})\hat{L}-\hat{n}}{(1-(\hat{L}\cdot\hat{n})^2)^{1/2}}, \frac{\hat{n}\times\hat{L}}{(1-(\hat{L}\cdot\hat{n})^2)^{1/2}}, \hat{L}\Bigg).
\end{equation}
where $\hat{L}$ is the angular momentum direction, $\gamma$ is the angle between the pericenter of the orbit and the $\hat{x}_{ \hat{L}}$. Transforming the quantity $I^{ij}$ further into the $\hat{n}$-based coordinate,
\begin{equation}
(\hat{x},\hat{y},\hat{z})_{\rm AK}:=\Bigg(\frac{\hat{n}\times\hat{L}}{(1-(\hat{L}\cdot\hat{n})^2)^{1/2}}, \frac{\hat{L}-(\hat{L}\cdot\hat{n})\hat{n}}{(1-(\hat{L}\cdot\hat{n})^2)^{1/2}}, -\hat{n}\Bigg),
\end{equation}
then, with the expression of the projection operator $P_{ij}$, one can obtain the two polarization coefficients $h^+, h^\times$. 

The orbital angular momentum $\hat{L}$ is precessing along the central BH's spin direction $\hat{S}$, while the central BH's spin direction can be approximated as fixed and has angular $(\theta_K, \phi_K)$ relative to the line of sight. Knowing the angle $\lambda$ between $\hat{L}$ and $\hat{S}$ and the Lense-Thirring precession azimuthal angle $\alpha$, one can express $\hat{L}$ in terms of $\theta_K,\phi_K,\alpha$  and $\lambda$. For the pericenter precession angle $\gamma$, one can also re-express it  into an intrinsic parameter $\tilde{\gamma}$  and an extrinsic parameter $\beta$, where $\tilde{\gamma}$ is the angle between pericenter and $\hat{L}\times\hat{S}$, and $\beta$ is the angle between $\hat{L}\times\hat{S}$ and $\hat{x}_{ \hat{L}}$. Thus, to calculate the SWD waveform, we have to know the parameters including
\begin{equation}
( M, \mu, \nu, e, \tilde{\gamma},\Phi, \theta_S,\phi_S, \lambda, \alpha, \theta_K, \phi_K,  D_L).
\end{equation}
Among those parameters, there are five quantities evolving with time and following PN equations, which are $(\Phi, \nu, e, \alpha, \tilde\gamma)$ (\cite{Barack:2003fp, Blanchet:2001aw}). The evolution equations of parameters $\Phi, \nu, e$ include terms up to 3.5PN order and $\alpha, \tilde{\gamma}$ include terms up to 2PN order. These equations related to quantities $ M, \mu, a, \nu_0, e_0, \Phi_0,\alpha_0$ and $\tilde{\gamma}_0$, where  $a$ is the central BH spin, $(  \nu_0, e_0,\Phi_0,\alpha_0$ and $\tilde{\gamma}_0)$ is the initial value of ($ \nu, e, \Phi, \alpha, \tilde{\gamma}$). Thus, we have 14 parameters to obtain the SWD waveform,
\begin{equation}
( M, \mu, a, \nu_0, e_0, \tilde{\gamma}_0,\Phi_0, \theta_S,\phi_S, \lambda, \alpha_0, \theta_K, \phi_K,  D_L).
\end{equation}

\subsection{Mass Loss}\label{MassLoss}
%WD tightly bound to the MBH will undergo a multiple-passage of mass transfer. The lossed  mass will be accreted by the MBH and generate electromagnetic signals. Zalamea+2010 shows that this tidal stripping occurs only near the pericenter of the orbit, even with a very small orbit eccentricity.  
%\citet
The WD will be tidally stripped upon each pericenter passage after arriving at the position of $R_{\rm strip}$, which would cause the mass loss  of the WD and enlarge the tidal disrupt radius. In this paper, we try to add the mass loss effect to the SWD waveform calculation.  \cite{Chen:2022oin} did a calculation of the amount of the  stripping mass when the WD orbit pericenter $R_{\rm peri}$ located between $0.5<\beta<0.7$, and obtain the stripping loss mass equation as
\begin{equation}\label{mloss}
\frac{\Delta{M}}{\mu}\simeq4.8\big[1-(\mu/M_{\rm ch})^{4/3}\big]^{3/4}\Big(1-\frac{\beta_0}{\beta}\Big)^{5/2},
\end{equation}
where $\beta_0\simeq0.5$. In principle, the GW signal is expected to  continue radiated when $0.7\leq\beta\leq1$. However, the simulation of the system shows that the WD will lose mass quickly in this period (\cite{Chen:2022oin}), reach the tidal disrupt radius in a very short time and end the GW signal. As mass lost in this period is quite uncertain by now and the GW signal is expected very short, we choose $\rm{max}(R_{\rm LSO}, R_{\rm p|\beta=0.7})$ to truncate the SWD waveform.

We add the mass loss to the WD and assume the WD ejects its mass immediately each time it arrives at the pericenter position, resulting in a decrease in the mass of $\Delta M$ based on equation (8). We expect the stripping mass will remain in the original orbit until accreted by the central BH slowly. Thus the unit mass of the remaining WD will not lose energy and angular momentum. 
 
%We expect that when the white dwarf (WD) reaches the pericenter position, it will immediately eject its lost mass, resulting in a decrease in mass of ΔM. We anticipate that the stripped mass will remain in its original orbit until it is slowly accreted by the central black hole (BH).

 %Does the mass get ejected and the WD becomes less massive as a result
 
 %The expected result is that the amplitude of the waveform will decrease and the WD will disrupt more quickly compared with the QPE waveform assuming no mass loss.
 
 %We give an assessment of this effect and showed it in section \ref{TD}. 
%We get the orbit pericenter according to the WD's orbital evolution equations and add the mass loss to the WD each time it arrives at the pericenter position. 

%For EMRI, the waveform will be cut off at the last stable orbit of the central massive black hole. While for QPE, the WD will be disrupt at the tidal disrupt radius
%\begin{equation}
%R_{\rm T}=R_*(M/\mu)^{1/3},
%\end{equation}
%and the GW signal will end. Here, $M, \mu$ are the MBH's mass  and the WD's mass respectively, $R_*$ is the WD's radius and has mass-radius relation
%\begin{equation}
%R_*=9\times10^8\Bigg[1-\Big(\frac{\mu}{M_{\rm ch}}\Big)^{4/3}\Bigg]^{1/2}\Big(\frac{\mu}{M_\odot}\Big)^{-1/3}\rm cm,
%\end{equation}
%where $M_{\rm ch}\simeq1.44M_\odot$ is the Chandrasekhar mass. 

\subsection{Tidal Deformability}\label{TD}
Besides the above effect, the WD approaching the central BH gets tidally deformed will change its multipolar structure, which may leave an imprint on the SWD waveform. So far, the impact of the tidal deformability on the GW signal  has been widely studied in the case of comparable masses (\cite{Valsecchi:2011mv,Flanagan:2007ix}). For the double white dwarfs, ignoring the tidal deformability effect on the GW signals will lead to a bias in the mass determination and misidentify WDs as neutron stars or black holes (\cite{Valsecchi:2011mv}). SWD has a much higher mass ratio, the tidal deformability effect has not been discussed sufficiently and should be further explored.  

%In this work, we try to give an assessment of the WD tidal deformability effect on the QPE waveform. 

\cite{Valsecchi:2011mv} investigated the pericenter precession of  a binary system, which consists of two stars with mass  $M_{1,2}$,  due to quadrupole tides in their work. Applying their equation into SWD systems,
the tidal deformability of the WD will lead to a precession rate of
\begin{equation}
\dot{\gamma}_{{\rm Tid}}=30\pi\nu\Big(\frac{R_*}{r_{\rm semi}}\Big)^5\frac{M}{\mu}\frac{1+\frac{3}{2}e^2+\frac{1}{8}e^4}{(1-e^2)^5}k_2,
\end{equation}
where $r_{\rm semi}$ is the orbital semi-major axis and has a relation $r_{\rm semi}=(2\pi M \nu)^{-2/3}M$, $k_2$ is the tidal Love number, which is used to quantify the deformability of the WD (\cite{Hinderer:2007mb}). The value of $k_2$ relies on the detailed WD structure and has expression (\cite{Valsecchi:2011mv})
\begin{equation}
k_2=\frac{1}{2}\Big(\frac{\xi(R_*, T)}{R_*}-1\Big),
\end{equation}
where $\xi$ is the radial component of the tidal displacement of the WD, $T$ is the temperature.
In the study of \cite{Deloye:2007uu}, a full structure exploration of the WD donors in binary evolution systems is presented, which include three phases: the mass transfer turn-on phase, the expanding phase in response to mass loss, and the cooling contraction phase.  Building on this, \cite{Valsecchi:2011mv} investigated the dimensionless quantity $k_2$  at different temperatures. Further studies are needed to determine the evolution value of $k_2$ in SWD systems. Here, we choose a much lower value of $k_2=0.014$, a medium value of $k_2=0.081$, and a relatively larger value of $k_2=0.15$ to assess the WD tidal deformability effect. 

%We'd like to provide more understanding of this effect on SWD systems.

%expect the WD will have a significant deformation after reaching $R_{\rm strip}$. 
%During the  detailed WD structure models from Deloye et al. (2007) to compute ki at different evolutionary phases during the WD lifetime.

%\cite{Valsecchi:2011mv} also provided an exploration of the dimensionless quantity $k_2$  at different temperatures.

We incorporated the tidal precession rate into the WD's orbital evolution equation and investigated the tidal phase correction to the GW signals.

\section{TianQin Gravitational Wave Detector}\label{detector}
TianQin is a geocentric space-based GW detector with three satellites forming a triangular constellation (\cite{TianQin:2015yph}). The direction of the constellation points to a white dwarf binary system RX J0806.3+1537 (short for J0806). TianQin has three arms and each armlength is about $1.7\times10^8$m. TianQin will operate for 5 yrs. Due to the heat instability, the observation scheme of TianQin is ``three months on+3 months off''.

A GW signal entering into the TianQin detector will have the response signal as (\cite{Rubbo:2003ap,Cornish:2002rt})
\begin{equation}
h(t)=F^+(t)h^+(t)+F^\times(t)h^\times(t),
\end{equation}
where $h^+, h^\times$ are the two independent polarization states of the GW signal, $F^+, F^\times$ are the antenna beam pattern factors, which are given by
\begin{equation}
F^+(t)=\frac{1}{2}[\cos(2\psi)D^+(t)-\sin(2\psi)D^\times(t)],
\end{equation}
where $\psi$ is the polarization angle. In the low frequency limit, $D^+, D^\times$ have the expressions as
\begin{equation}
\begin{aligned}
D^+(t)=&[\hat{r}_{12}(t)\otimes\hat{r}_{12}(t)-\hat{r}_{13}\otimes\hat{r}_{13}(t)]:{\mathbf e}^+,\\
D^\times(t)=&[\hat{r}_{12}(t)\otimes\hat{r}_{12}(t)-\hat{r}_{13}\otimes\hat{r}_{13}(t)]:{\mathbf e}^\times.
\end{aligned}
\end{equation}
$\hat{r}_{ij}$ is the unit direction vector between satellite $i$ and satellite $j$ with equation 
 \begin{equation}
\hat{r}_{ij}= (\mathbf{x}_j-\mathbf{x}_i)/L_{ij}, 
 \end{equation}
where ``:” is double contraction,  $\mathbf{x}_i$ is the  position of the satellite $i$ in the ecliptic coordinate, which can be found in \cite{Fan:2020zhy}, $L_{ij}$ is the armlength between satellite $i$ and satellite $j$, which we assume will remain constant. $\mathbf{e}^+$ and $\mathbf{e}^\times$ are two basis tensors in the $\hat{n}$-based coordinate
\begin{equation}
\begin{split}
\mathbf{e}^+&=\hat{u}\otimes\hat{u}-\hat{v}\otimes\hat{v},\\
\mathbf{e}^\times&=\hat{u}\otimes\hat{v}+\hat{v}\otimes\hat{u},
\end{split}
\end{equation}
where $\hat{u}$ and $\hat{v}$ are
\begin{equation}
\begin{split}
\hat{u}&=(\cos\theta_S\cos\phi_S, \cos\theta_S\sin\phi_S, -\sin\theta_S),\\
\hat{v}&=(\sin\phi_S, -\cos\phi_S).
\end{split}
\end{equation}

TianQin is aiming to detect GW sources in the frequency band $10^{-4}-1$Hz. The noise model of TianQin is encoded in the following sensitivity curve
\begin{equation}
S_n(f)=\frac{1}{L^2}\Big[\frac{4S_a}{(2\pi{f})^4}\big(1+\frac{10^{-4}\rm Hz}{f}\big)+S_x\Big]\times\Big[1+0.6\big(\frac{f}{f_*}\big)^2\Big],
\end{equation}
where $S_a^{1/2}=1\times10^{-15}\rm m \rm s^{-2}/\rm Hz^{1/2}$ and $S_x^{1/2}=1\times10^{-12}\rm m/\rm Hz^{1/2}$ are the residual acceleration noise and position noise, respectively. $f_*=1/(2\pi{L})$ is the transfer frequency.

%The  In Sec.\ref{MassLoss} and Sec.\ref{TD}, we mentioned that the SWD sources will ,
\section{Method and Result}\label{result}
\subsection{Waveform Mismatch}
%Except for being specific highly eccentric and relatively lower mass EMRI sources, SWDs will experience mass loss and tidal deformability, as we mentioned Since the detailed WD structure models during the evolution have a lot of uncertainty, we only provide a preliminary exploration to determine whether the effects can be ignored for SWD sources.
In Sec.\ref{MassLoss} and Sec.\ref{TD}, we mentioned that the SWD sources will experience mass loss and tidal deformability, which may influence their waveforms.  Here, we use a function called fitting factor to quantify the waveform mismatch induced by these two effects. Defining the noise-weighted inner product between two signals $s_1(t)$ and $s_2(t)$ as (\cite{Finn:1992wt})

\begin{equation}
(s_1|s_2)=2\int^\infty_0\frac{\tilde{s}_1(f)\tilde{s}^*_2(f)+\tilde{s}_2(f)\tilde{s}^*_1(f)}{S_n(f)}{\rm d}f,
\end{equation}
where $\tilde{s}(f)$ is the Fourier transforms of $s(t)$. The fitting factor has expression as (\cite{Creighton2011GravitationalwavePA})

\begin{equation}\label{ff}
FF=\frac{(h(\theta)|h^\prime(\theta))}{\sqrt{(h(\theta)|h(\theta))(h^\prime(\theta)|h^\prime(\theta))}},
\end{equation}
where $h(\theta)$ is the original waveform, $h^\prime(\theta)$ is the influenced waveform.  

 \begin{figure}
 \flushleft
%\centering
%\includegraphics[width=\columnwidth,clip=true,angle=0,scale=0.75]{Figure/SNRdistri.pdf}
\includegraphics[width=\linewidth]{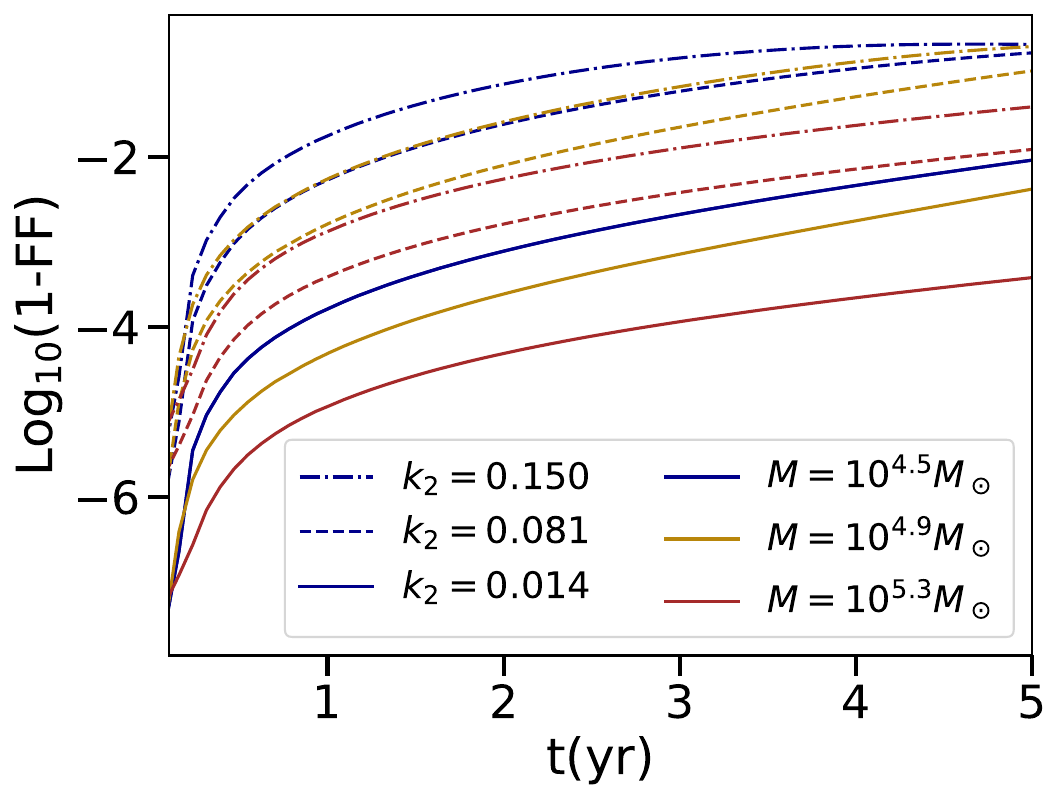}
\caption{The waveform mismatch under the effect of tidal deformability, with WD mass $\mu=0.6M_\odot$. The solid lines describe the evolution of the fitting factor with $k_2 = 0.014$, the dashed lines describe the evolution of the fitting factor
with $k_2 = 0.081$, and the dash-dot lines describe the evolution of the fitting factor with $k_2 = 0.15$. The blue lines correspond to SWD sources with $M=10^{4.5}M_\odot$, the yellow lines correspond to SWD sources with $M=10^{4.9}M_\odot$, and the red lines correspond to SWD sources with $M=10^{5.3}M_\odot$. }
\label{fig:fittingFac}
\end{figure}

We generate about one thousand SWD sources, with $M, \mu$ drawn among the blue shadow region of Fig.\ref{fig:eventdistribut}, $D=10 \rm kpc$, $e_0=0.9$, the MBH spin are drawn uniformly from $(0, 1)$, the sky position $(\theta_S, \phi_S)$ and the spin orientation $\theta_K, \phi_K$ are drawn from an isotropic distribution on the sphere, $\lambda$ are drawn from $(0, \pi)$, $\tilde{\gamma}_0, \Phi_0, \alpha_0$ are drawn from $(0, 2\pi)$, $\nu_0$ are drawn from $10^{-4}\sim10^{-3}$Hz, and the observation time is 5yrs.  Then, three groups of SWD signals are generated without any effect, with mass loss, with tidal deformability,  respectively. 

The result shows that the mass loss will shorten the SWD waveform and decrease its amplitude, as the smaller the WD mass is, the larger the tidal disrupt radius and the weaker the signal strength. As for the radiated waveform, the fitting factor approaches 1. This is because the evolution of the WD orbit depends on the energy, the angular momentum, and the Carter constant. In our study, we assume the stripping mass will remain in its original orbit until it is accreted by the central BH slowly. As a result, there will be no change in the energy, angular momentum, and the Carter constant for the remaining unit WD mass. Consequently, the orbit of the WD will remain unchanged and the waveform phase will not be affected.

%the evolution of the WD orbit depends on three quantities: energy, angular momentum, and the Carter constant.

%has little influence on the phase of the gravitational waveforms. ,  as the smaller the WD mass is, the larger the tidal disrupt radius

As for the tidal deformability, the result shows that the impact of the tidal deformability on the SWD waveform is relatively noticeable.  We selected three SWD sources and show the result in Fig.\ref{fig:fittingFac}. In this figure,  lines describe the evolution of the fitting factor with $k_2=0.014$, dashed lines describe the evolution of the fitting factor with $k_2=0.081$, and dash-dot lines describe the evolution of the fitting factor with $k_2=0.15$. From this figure, we can find that the waveform mismatch is relatively small at the beginning of the waveform. With the increase of the WD evolution time, the waveform mismatch increases.

Compared with mass loss, the tidal deformability of the WD will lead to a pericenter precession, which will leave an imprint on the phase of the SWD waveform. The impact of the tidal deformability on the SWD waveform with $k_2=0.014$ can be negligible, as the waveform mismatch is lower than $10^{-3}$ in the above figure. When considering the impact of the tidal deformability with $k_2=0.081$ and $k_2=0.15$, the finding  reveals that the waveform mismatch for specific SWD sources can increase to 0.2 over time. 

We further investigated the impact of the tidal deformability on the maximum detection distance and the intrinsic parameter estimation precision of the SWD signals, with the method detailed in Sec.\ref{Sec:horiDis} and Sec.\ref{Sec:ParaEsti}. We presented the results in Table \ref{tab:maxD} and Table \ref{tab:ParaEsti}, where the SWD sources are in accordance with the three sources listed in Fig.\ref{fig:fittingFac}.  The results show that the impact of the tidal deformability on the maximum detection distance and the intrinsic parameter estimation precision can be ignored. This is evident as the number of wave cycles has little change. Furthermore, this seemingly unfavorable scenario in the fitting factor and the highly accurate parameter estimation remind us that an inaccurate waveform could potentially result in the misidentification of parameters for SWD sources, which highlights the need for further comprehensive exploration in the future. 

%Such a seemingly unfavorable scenario

\begin{table}
\caption{The influence of the tidal deformability on the maximum detection distance, with the SWD sources same as the three sources listed in Fig.\ref{fig:fittingFac}.}\label{tab:maxD}
\begin{center}
\setlength{\tabcolsep}{1.6mm}
%	\resizebox{\columnwidth}{!}{%
\begin{tabular}{|cc|c|c|c|c|}
\hline
&BH mass($M_\odot$)&$k_2$=0&$k_2$=0.150&$k_2$=0.081&$k_2$=0.014\\\hline
\multirow{3}{*}{D(Mpc)}&log$_{10}{M} = 4.5$&31.04&31.04&31.04&31.04\\
&log$_{10}{M} = 4.9$&66.79&66.80&66.79&66.79\\
&log$_{10}{M} = 5.3$&102.86&102.81&102.84&102.86\\
\hline
\end{tabular}
\end{center}
\end{table}

\begin{table}
\caption{The influence of the tidal deformability on the intrinsic parameter estimation precision, with the SWD sources same as the three sources listed in Fig.\ref{fig:fittingFac}.}\label{tab:ParaEsti}
\begin{center}
\setlength{\tabcolsep}{1.6mm}
%	\resizebox{\columnwidth}{!}{%
\begin{tabular}{|cc|c|c|c|c|}
\hline
&BH mass($M_\odot$)&$k_2$=0&$k_2$=0.150&$k_2$=0.081&$k_2$=0.014\\\hline
\multirow{3}{*}{$\frac{\Delta M}{M}$}&log$_{10}{M} = 4.5$&1.291e-5&1.291e-5&1.291e-5&1.291e-5\\
&log$_{10}{M} = 4.9$&1.683e-6&1.683e-6&1.683e-6&1.683e-6\\
&log$_{10}{M} = 5.3$&4.808e-7&4.814e-7&4.810e-7&4.809e-7\\\hline
\multirow{3}{*}{$\frac{\Delta m}{m}$}&log$_{10}{M} = 4.5$&1.024e-6&1.024e-6&1.024e-6&1.024e-6\\
&log$_{10}{M} = 4.9$&2.050e-6&2.050e-6&2.050e-6&2.050e-6\\
&log$_{10}{M} = 5.3$&7.412e-7&7.418e-7&7.413e-7&7.412e-7\\\hline
\multirow{3}{*}{$a$}&log$_{10}{M} = 4.5$&2.166e-5&2.166e-5&2.166e-5&2.166e-5\\
&log$_{10}{M} = 4.9$&5.580e-6&5.581e-6&5.580e-6&5.580e-6\\
&log$_{10}{M} = 5.3$&3.238e-6&3.255e-6&3.244e-6&3.238e-6\\\hline
\multirow{3}{*}{$e_0$}&log$_{10}{M} = 4.5$&5.231e-8&5.231e-8&5.231e-8&5.231e-8\\
&log$_{10}{M} = 4.9$&1.501e-8&1.501e-8&1.501e-8&1.501e-8\\
&log$_{10}{M} = 5.3$&5.143e-8&5.148e-8&5.144e-8&5.143e-8\\\hline
\end{tabular}
\end{center}
\end{table}

%\multirow{3}{*}{SNR}&log$_{10}{M} = 4.5$&55899.75&120268.26&25(11)&8 (1)\\
%&log$_{10}{M} = 4.9$&120268.26&120268.26&25(11)&8 (1)\\
%&log$_{10}{M} = 5.3$&185222.86&120268.26&25(11)&8 (1)\\\hline
%Our result shows that the larger the central BH and the smaller the WD, along with a lower initial frequency, the  fitting factor is closer to 1. Compared to the stripped mass, the impact of the tidal deformability on the QPE waveform is relatively larger.

%We choose a QPE source with parameters $M=10^5M_\odot$, $\mu=0.6M_\odot$, $\nu_0=1.0e^{-4}$

\subsection{Horizon Distance}\label{Sec:horiDis}
 TianQin's capability of detecting SWD can be assessed by the horizon distance, which is the farthest distance that a SWD can be detected.  That is, assuming those SWDs are at the most favorable detectable conditions relative to the TianQin detector,  the horizon distance is those sources at which the SNR exceeds the detection threshold.

To determine the detection threshold for SWD, we follow the previous studies on EMRI, which set the threshold value as 15 (\cite{MockLISADataChallengeTaskForce:2009wir}). For one Michelson interferometer, the SNR calculation (\cite{Finn:1992wt}) is defined as
\begin{equation}
\rho=(h|h)^{1/2}=2\Big[\int^\infty_0\frac{\tilde{h}(f)\tilde{h}^*(f)}{S_n(f)}{\rm d}f\Big]^{1/2},
\end{equation}
where $h$ is the response GW signal, $S_n(f)$ is the power spectral density of TianQin detector.

TianQin has three arms, constructing two independent interferometers. The SNR calculation has an expression as
\begin{equation}
\rho=\sqrt{\rho^2_I+\rho^2_{II}}.
\end{equation}

\begin{figure}
\centering
\includegraphics[width=\linewidth]{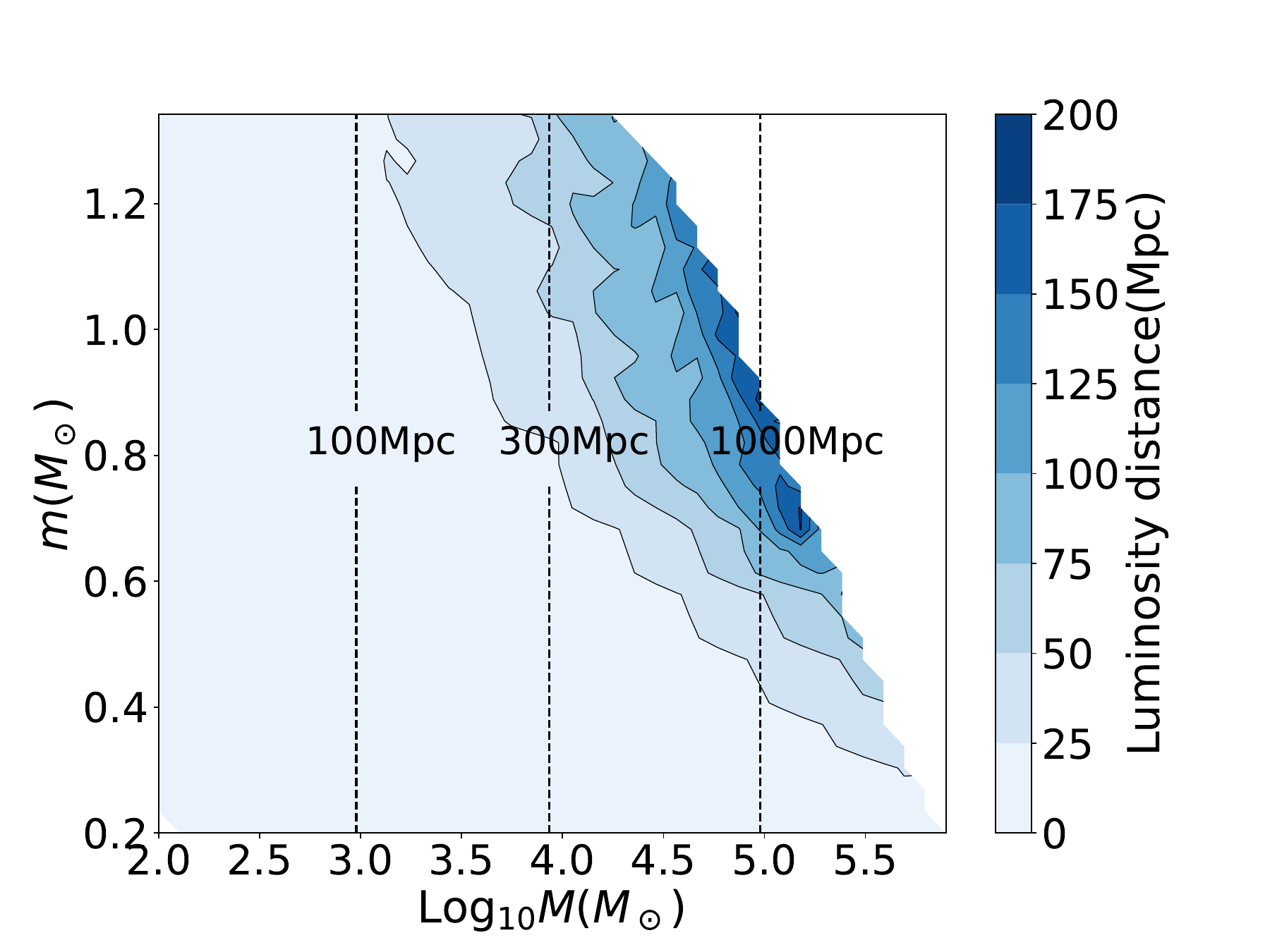}
%\caption{The dark red line corresponds to the EMRB events number distribution with SNR above 4 and the shadow corresponds to the uncertainty assuming a Poisson distribution. }
\caption{The blue shadow region represents the GW horizon distance of TianQin for SWD GW systems. The dark dashed lines are the maximum detection distance for the electromagnetic emission from the horizon distance for SWDs using the Einstein Probe. The luminosity of the electromagnetic emission is capped at the Eddington luminosity of the central BHs.}
\label{fig:snrdistribut}
\end{figure}

We try to obtain the horizon distance for the SWD sources within the blue shadow region of Fig.\ref{fig:eventdistribut}. We set  the initial value of $\phi, \tilde{\gamma}, \alpha$ to be 0, as they have little influence on the horizon distance. We fix the inclination angle $\lambda$ to $\pi/2$, and $\theta_K, \phi_K$ to $\pi/4$. We assume the central BHs have spin values of $\sim0.98$. Furthermore, the space positions of these sources are placed in the direction of J0806, which is the orientation of the TianQin detector.  
The initial frequency $\nu_0$ has a significant impact on the SWD sources. 
A higher initial frequency  indicates that the WD is closer to the central BH, resulting in a stronger waveform and a shorter evolution time. On the other hand,  a lower initial frequency  implies that the WD is far away from the central BH, leading to a relatively weaker waveform and a longer evolution time. 
To determine the optimal frequency value $\nu_0$, we drew it uniformly between $10^{-4}$ and the frequency at the last orbit, We then calculate the SNR for each source and select the one with the largest SNR as the optimal frequency source. The SNR is inversely proportional to the luminosity distance. Using this information, we  can determine the horizon distance. The calculation of the horizon distance is based on the optimal parameters, which does not imply that the BHs that are sources of SWDs all have these parameter values.

%The close the CO is to the MBH, the higher frequency and the larger amplitude of the QPE GW is. While the far away the CO from the MBH, the lower frequency and the smaller amplitude of the QPE GW is. At the same time, the further the CO from the central BH, the longer the evolution time and the easier it is to accumulate a higher SNR.
%After adding the mass loss and the tidal deformability effects to the QPE waveform,

 We show the result  in Fig.\ref{fig:snrdistribut}. In this figure, the shadow region represents the horizon distance, while the blank area is because there has no source. From this figure, we can find that for most SWD sources, the horizon distance does not exceed 100Mpc, and no SWD sources above 200Mpc are expected to be detected by TianQin. Therefore, we can set 200Mpc as the horizon distance for all the SWD sources. From this figure, we can also find that the main detectable SWD signals for TianQin are produced by central BHs with masses between $10^4 \sim 10^{5.5}M_\odot$, and the maximum horizon distance corresponds to the central BH with a mass of $M=10^5M_\odot$. This phenomenon  is related to the sensitivity curve of TianQin and the characteristics of SWD signals, as with the increase of the central BH mass, the amplitude of GWs increases while the frequency decrease. 
%the more massive the WD is, the farther the horizon distance is.
%By now, five QPEs have been discovered so far, their parameters has been collectively shown in Chen et al

The SWD sources have both gravitational wave signals and electromagnetic signals, which can be used to perform cosmological inference. We expect the electromagnetic emission from SWD sources is bright in X-ray, and the luminosity is capped at the Eddington luminosity $L_{\rm Edd}$ of the black hole. Thus, the maximum detection distance can be roughly estimated by $d_{\rm EM} \simeq [L_{\rm Edd}/(4 \pi F_{\rm lim})]^{1/2}$. Here we adopt the sensitivity of the Follow-up X-ray Telescope (FXT) of the Einstein Probe \citep[EP, ][]{Yuan_EP_2015}, which is $F_{\rm lim} \simeq 10^{-13}\ {\rm erg\ s^{-1}\ cm^{-2}}$ in 0.5 -- 10 keV with $1$ ks exposure time \citep{Zhang_EPsensitivity_2022}. We can obtain that the maximum detection distance is $d_{\rm EM} \simeq 323(M/10^4\ M_{\odot})^{1/2}$ Mpc, which are the dashed lines shown in Fig. \ref{fig:snrdistribut}.

%seemingly
{
Notably, if SWD sources are hyper-Eddington events, with their hyper-Eddington energy radiation likely originating from relativistic jets \citep{ye_tidal_2023}, they can be observed at a larger distance than we estimated here. QPEs have been proposed as electromagnetic counterparts of SWDs, with their Eddington-limited X-ray luminosity likely stemming from the accretion disk. However, as of now, no jet emissions have been observed from QPEs, Nonetheless, we are looking forward to the discovery of a SWD accompanied by a jet.
}

\subsection{Parameter Estimation Precision}\label{Sec:ParaEsti}

\begin{figure*}
	\centering
	\begin{minipage}{0.45\linewidth}
		\centering
		\includegraphics[width=\linewidth]{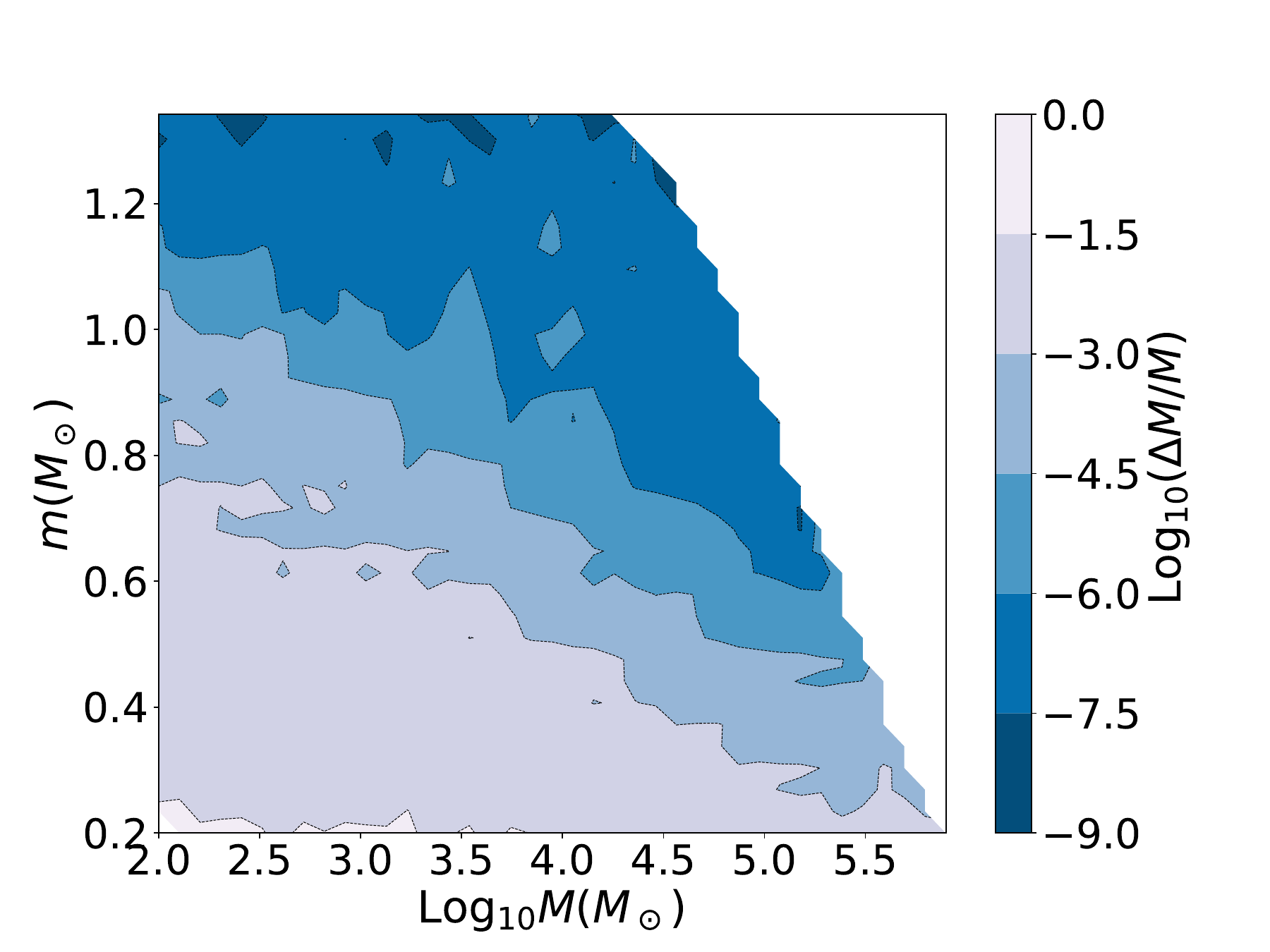}
		%\label{param}
	\end{minipage}
	\begin{minipage}{0.45\linewidth}
		\centering
		\includegraphics[width=\linewidth]{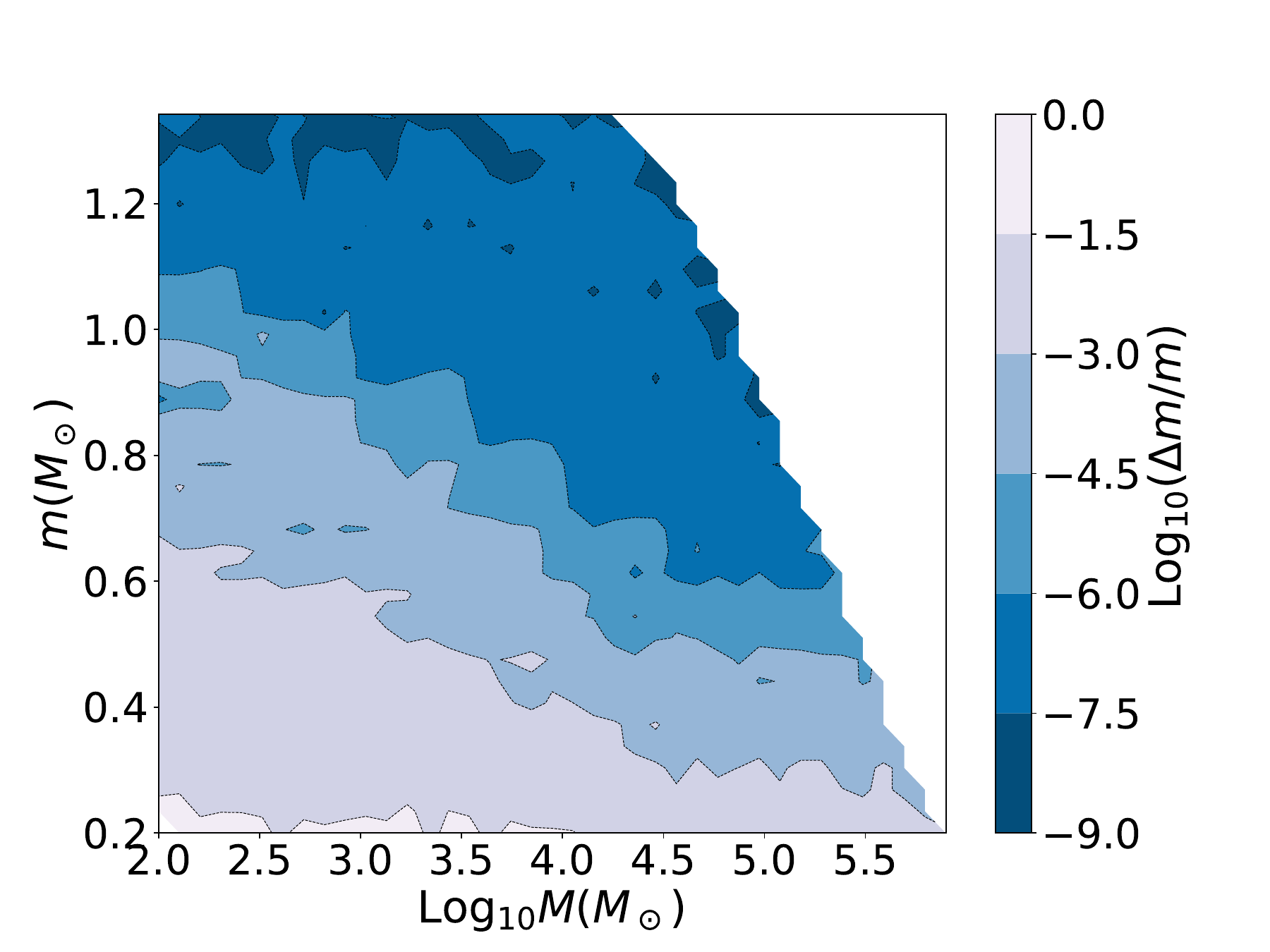}
	%	\label{paraa}
	\end{minipage}
		\begin{minipage}{0.45\linewidth}
		\centering
		\includegraphics[width=\linewidth]{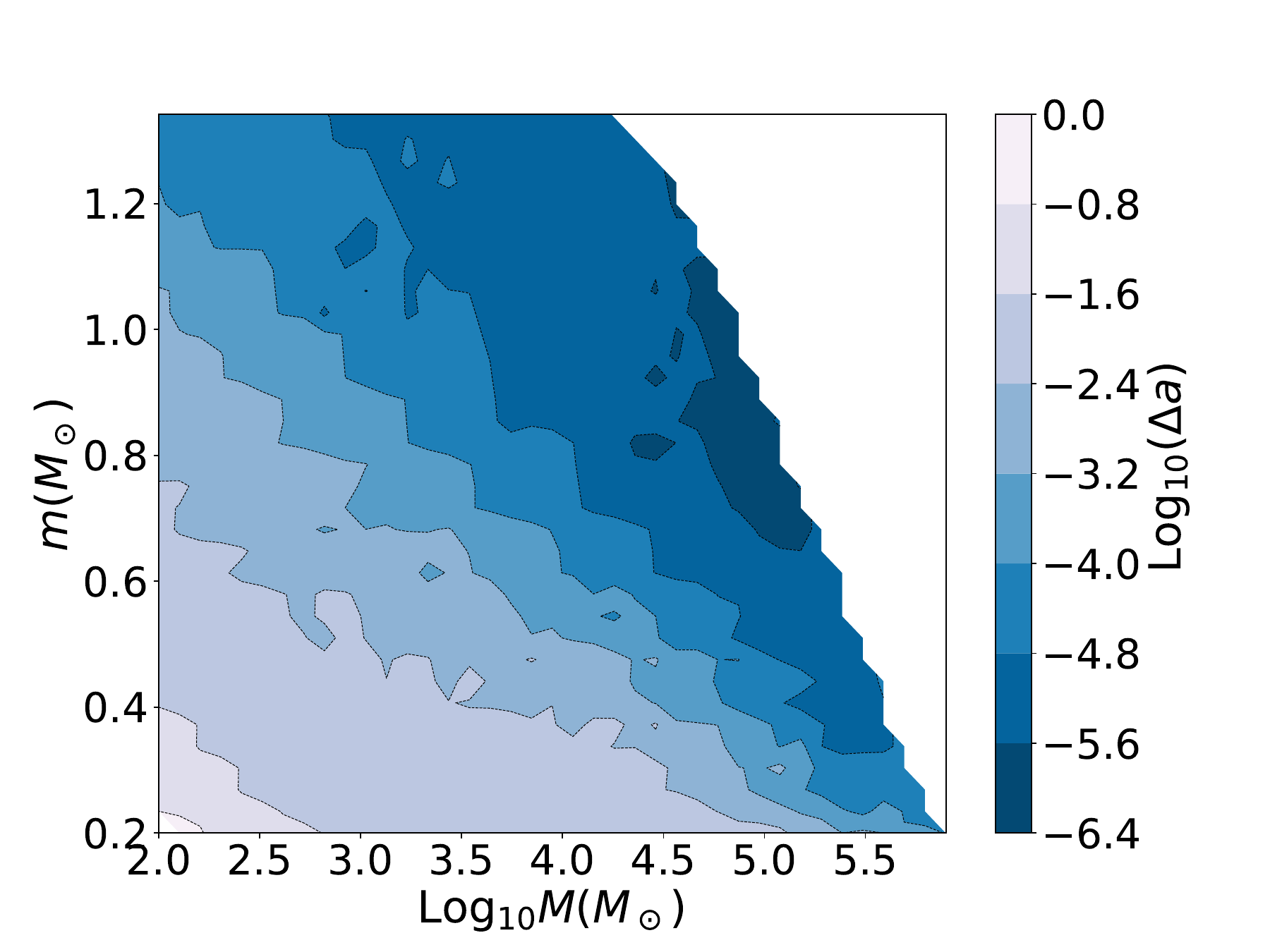}
		%\label{paramm}
	\end{minipage}
	\begin{minipage}{0.45\linewidth}
		\centering
		\includegraphics[width=\linewidth]{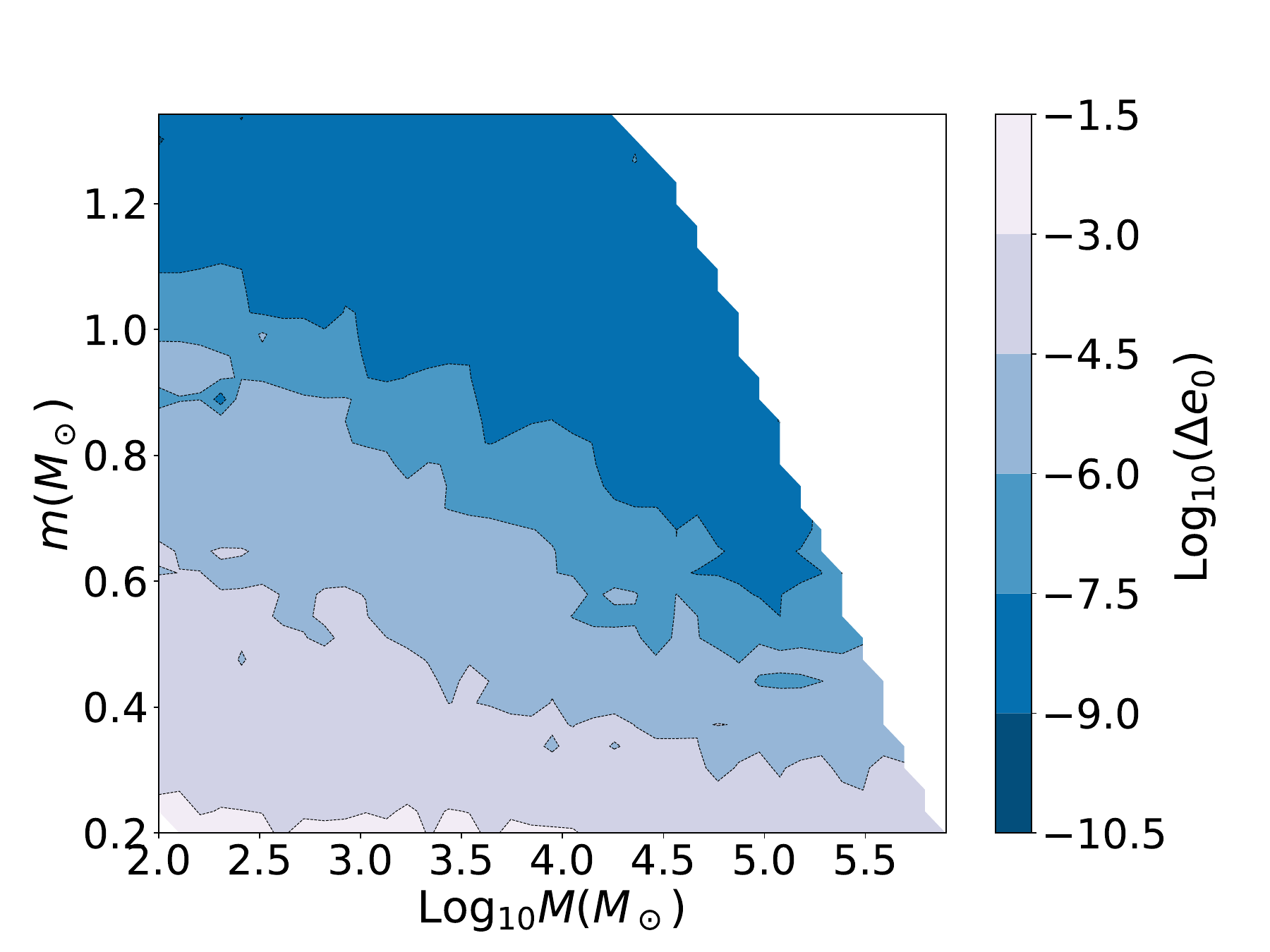}
		%\label{paraa}
	\end{minipage}
		\begin{minipage}{0.45\linewidth}
		\centering
		\includegraphics[width=\linewidth]{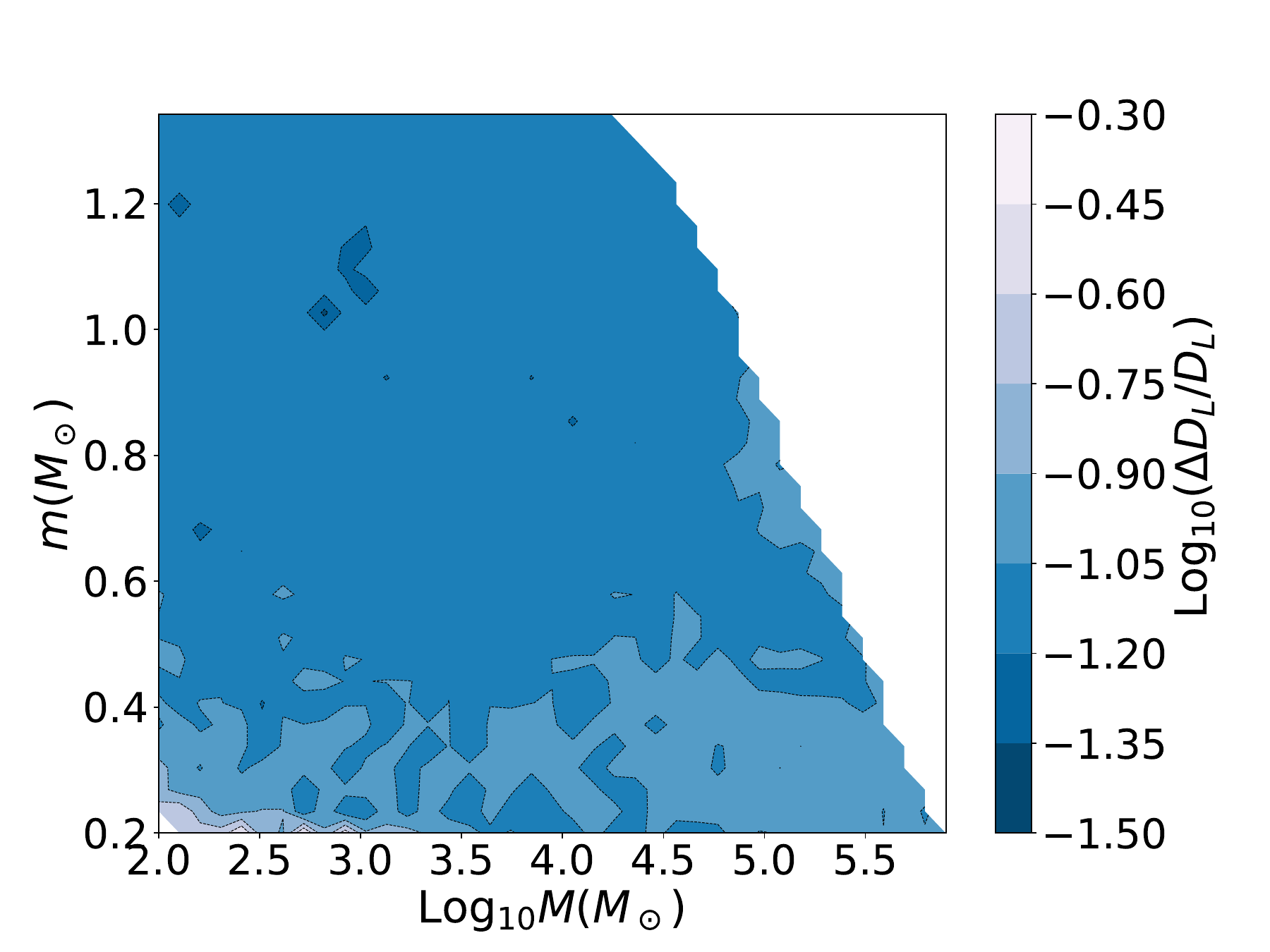}
		%\label{paraa}
	\end{minipage}
		\begin{minipage}{0.45\linewidth}
		\centering
		\includegraphics[width=\linewidth]{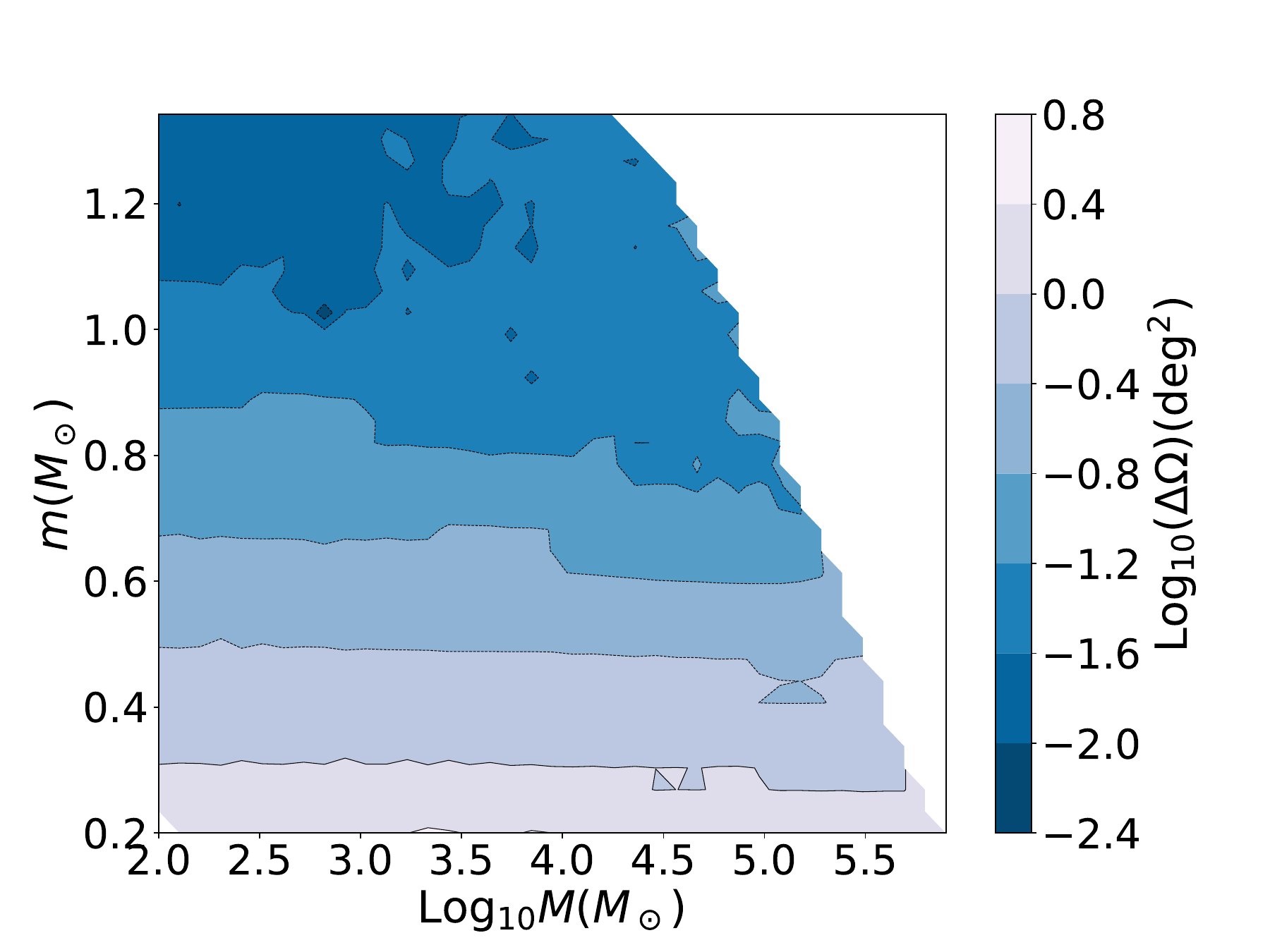}
		%\label{paraa}
	\end{minipage}
\caption{The parameter estimation precision for various parameters by TianQin. The upper plots show the result for $M$ and $\mu$, the middle plots display the result for $a$ and $e_0$ and the lower plots illustrate the result for $D_L$ and $\Omega$.}
\label{paraEsti}
\end{figure*}

The detection of  SWD sources can provide valuable information about celestial bodies, leading to a  better understanding of the universe. However, this relies on high-precision measurements of the source parameters. 

Assuming an SWD signal with true physical parameter $\Theta$, the existence of noise will lead to uncertainties in the inference of source parameters. One routine method is to use the Fisher information matrix (\cite{Vallisneri:2007ev,Rodriguez:2013mla}) to characterize the parameter estimation performance of the gravitation-wave measurements, which is an approximation on the statistical uncertainty and can be achieved in the linearized-signal approximation with high SNR. Defining the FIM as
\begin{equation}
\Gamma_{ij}=\Big(\frac{\partial\tilde{h}(f)}{\partial\Theta^i}\Big|\frac{\partial\tilde{h}(f)}{\partial\Theta^j}\Big).
\end{equation}
Then the Cramer-Rao bound of the covariance matrix can be obtained as 
\begin{equation}
\Sigma_{ij}=\langle\delta\Theta_i\delta\Theta_j\rangle=(\Gamma^{-1})_{ij}.
\end{equation}
Therefore, the estimation uncertainty for the $i$th parameter has the expression as
\begin{equation}
\sigma_i=\Sigma^{1/2}_{ii}.
\end{equation}
To express the sky localization uncertainty of the source, the solid angle $\Delta\Omega$ corresponding to the error ellipse can be obtained as
\begin{equation}
\Delta\Omega=2\pi|\sin\theta_S|\sqrt{\Sigma_{\theta_S}\Sigma_{\phi_S}-\Sigma^2_{\theta_S\phi_S}},
\end{equation}
where $\Sigma_{\theta_S}$ and  $\Sigma_{\phi_S}$ are the uncertainties on the ecliptic latitude angle $\theta_S$ and ecliptic longitude angle $\phi_S$, respectively.

For the SWD sources located at the horizon distance, as shown in Fig.\ref{fig:snrdistribut}, we give an assessment to investigate the parameter estimation precision of TianQin on these sources. These sources have the optimal conditions, corresponding to a relatively optimal parameter estimation precision. 
We present the result in Fig.\ref{paraEsti}. In this Figure, the shadow region is the parameter estimation precision for SWD source parameters. The upper plots show the result for $M$ and $\mu$, the middle plots display the result for $a$ and $e_0$ and the lower plots illustrate the result for $D_L$ and $\Omega$.

From this figure, we can find that the intrinsic parameters can be determined with precision much better than those extrinsic parameters. This is because the intrinsic parameters, such as the central BH mass $M$ and the WD mass $\mu$ and the spin $a$ of the central BH and the orbit initial eccentricity $e_0$, are related to the phase of the GWs. As the WD inspirals around the central BH for many cycles, even a slight deviation in its intrinsic parameters will  be clearly reflected in the GW phase. While the extrinsic parameter, such as the luminosity distance $D_L$ and the sky localization $\Omega$,  primarily affect the amplitude of the GWs, their estimation precision can't be improved by accumulating the phase mismatch over time.

From this figure, we also find that the estimation precision of the intrinsic parameters is widely distributed. We noticed that the parameters of SWD sources corresponding to a larger horizon distance and a massive WD in Fig.\ref{fig:snrdistribut} are more accurately estimated. This is because those SWD sources 
 have WDs evolving more rapidly around the central BH, making them more sensitive to phase mismatches.  While for those SWD sources corresponding to a smaller horizon distance and a small WD, the evolution is relatively minor, which will not contribute much to the phase mismatch. However, even in the worst-case scenario, TianQin can determine the central BH mass, the WD mass, the central BH spin, and the initial eccentricity with a precision of $10^{-2}$.  In a relatively optimistic case, the central BH mass and the WD mass can be determined with a precision of $10^{-7}$, the central BH spin can be determined with a precision of $10^{-5}$ and the eccentricity can be determined with a precision of $10^{-8}$. 

The estimation precision of the extrinsic parameters is relatively narrow. We expect TianQin can determine the luminosity distance with a precision of $10^{-1}$, and  determine the sky localization with a precision of $10^{-2}\sim10$ $\rm deg^2$. This result is obvious for the luminosity distance, as they have the same SNR. The sky location is associated with the Doppler phase shift.  As those SWD sources with higher WDs evolve more rapidly in frequency, their sky locations are more sensitive to parameter changes and  can be better estimated.

%To be more generally, we expect these intrinsic parameters can be determined with a precision of $10^{-5}$. 

\section{Conclusions and Discussion}\label{conclusions}

In this study, we focus on SWD detection with TianQin. We described the SWD source distribution based on the tidal stripping radius  and the LSO of the WD. We utilized AK to calculate the SWD waveform and applied a higher-order mode in the waveform calculation as SWD has a quite eccentric orbit. We also investigate the effect of mass loss and tidal deformability on the SWD waveform. Finally, we conduct a investigation into the horizon distance and  parameter estimation precision for SWD sources with TianQin.

Our result shows that the mass loss will shorten the SWD waveform and decrease its amplitude, but have little influence on the phase of the waveform. Compared with the mass loss, tidal deformability has a relatively noticeable effect on the waveform phase. 

As for the horizon distance, our result shows that by choosing a detection threshold of 15,  the horizon distance of SWD can be set to 200Mpc.  Moreover, we expect those SWD sources with central BH masses within $10^4\sim10^{5.5}M_\odot$ are more likely to be detected by TianQin. We calculate the maximum detection distance for the electromagnetic emission from SWD sources using the Einstein Probe, the result shows that the electromagnetic emission has a larger horizon distance than the GWs.

Our assessment of the estimation precision of TianQin on these SWD source parameters shows that, in the worst case, TianQin can determine the central BH mass, the WD mass, the central BH spin, and the initial eccentricity with a precision of $10^{-2}$. In the optimistic case, TianQin can determine the central BH mass and the WD mass with a precision of $10^{-7}$, determine the central BH spin with a precision of $10^{-5}$, and determine the eccentricity with a precision of $10^{-8}$.  Furthermore, TianQin can determine the luminosity distance with a precision of $10^{-1}$, and determine the sky localization with a precision of $10^{-2}\sim10$ $\rm deg^2$. 

The result presented above is preliminary. Especially, during the waveform calculation, we assume the structure of the SWD sources stays unchanged, with $k_2$ being constant, which does not reflect the real evolutionary process. More realistic evolution values of $k_2$ for SWD systems also require further studies. Additionally, the parameter estimation corresponds to a relatively optimal precision, as those sources have the optimal conditions such as the spin equal to 0.98, and the sky position located at J0806. An astrophysical model that capture the SWD populations is necessary to provide more realistic information on the parameter estimation precision of TianQin for SWD signals.

Both TianQin and LISA will be able to detect SWD sources but up to different distances and parameter accuracy (\cite{Fragione:2022ams, Torres-Orjuela:2023hfd}). A network of these two detectors could open up the possibility of studying a broader parameter space and enhance the astronomical information we can obtain. Consequently, a combined study of TianQin+LISA on SWD sources would be an interesting and rewarding work in the future.

\section*{Acknowledgements}
We are grateful to Rong-Feng Shen for his helpful discussion and advice. This work has been supported by Guangdong Major Project of Basic and Applied Basic Research (Grant No. 2019B030302001), the Natural Science Foundation of China (Grant No. 12173104), the Guangdong Basic and Applied Basic Research Foundation (Grant No. 2023A5150301).

\section*{Data Availibility}
No new data were generated or analysed in support of this research.
%
%%%%%%%%%%%%%%%%%%%%%%%%%%%%%%%%%%%%%%%%%%%%%%%%%%
%\section*{Data Availability}

%The underlying SkyMapper and ANU 2.3m/WiFeS data will be made available on reasonable request to the authors.



\bibliographystyle{mnras}
\bibliography{reference} % if your bibtex file is called example.bib

%\begin{thebibliography}{99}
%\end{thebibliography}

%%%%%%%%%%%%%%%%%%%%%%%%%%%%%%%%%%%%%%%%%%%%%%%%%%

%%%%%%%%%%%%%%%%% APPENDICES %%%%%%%%%%%%%%%%%%%%%

%%%%%%%%%%%%%%%%%%%%%%%%%%%%%%%%%%%%%%%%%%%%%%%%%%

% Don't change these lines
\bsp	% typesetting comment
\label{lastpage}
\end{document}